\documentclass[journal]{IEEEtran}

\usepackage{ifpdf}
\usepackage{authblk}

\usepackage{cite}

\ifCLASSINFOpdf
  \usepackage[pdftex]{graphicx}
\else
  \usepackage[dvips]{graphicx}
\fi

\usepackage{amsmath,bm}
\usepackage{amsfonts}
\usepackage{amssymb}
\usepackage{color,soul}
\usepackage{soul,color}
\soulregister\cite7
\soulregister\ref7

\usepackage{algorithmic}

\usepackage{array}

\usepackage{fixltx2e}

\newcommand\blfootnote[1]{%
  \begingroup
  \renewcommand\thefootnote{}\footnote{#1}%
  \addtocounter{footnote}{-1}%
  \endgroup
}

\begin{document}
\title{Recurrent Neural Networks For Accurate RSSI Indoor Localization}
\author{Minh Tu Hoang, Brosnan Yuen, Xiaodai Dong, Tao Lu, Robert Westendorp, and Kishore Reddy}
\maketitle

\begin{abstract}
\blfootnote{
This work was supported in part by the Natural Sciences and Engineering Research Council of Canada under Grant 520198, Fortinet Research under Contract 05484 and Nvidia under GPU Grant program (\textit{Corresponding authors: X. Dong and T. Lu.}). \\
M. T. Hoang, B.Yuen, X. Dong and T. Lu are with the
Department of Electrical and Computer Engineering, University of Victoria,
Victoria, BC, Canada (email: \{xdong, taolu\}@ece.uvic.ca).\\
R. Westendorp and K. Reddy are with Fortinet Canada Inc., Burnaby, BC,
Canada.}
This paper proposes recurrent neural networks (RNNs) for WiFi fingerprinting indoor localization. Instead of locating a mobile user's position one at a time as in the cases of conventional algorithms, our RNN solution aims at trajectory positioning and takes into account the correlation among the received signal strength indicator (RSSI) measurements in a trajectory. To enhance the accuracy among the temporal fluctuations of RSSI, a weighted average filter is proposed for both input RSSI data and sequential output locations. The results using different types of RNN including vanilla RNN, long short-term memory (LSTM), gated recurrent unit (GRU), bidirectional RNN (BiRNN), bidirectional LSTM (BiLSTM) and bidirectional GRU (BiGRU) are presented. On-site experiments demonstrate that the proposed structure achieves an average localization error of 0.75~m with $\mathrm{80\%}$ of the errors under one meter, which outperforms KNN algorithms and probabilistic algorithms by approximately $\mathrm{30\%}$ under the same test environment.  

\textit{Index Terms}- Received signal strength indicator, WiFi indoor localization, recurrent neuron network, long short-term memory, fingerprint-based localization. 
\end{abstract}

\section{Introduction} \label{sec:intro} 

The motivation of this paper is to locate a walking human using the WiFi signals of the carried smartphone. In general, there are two main groups in WiFi indoor localization: model based and fingerprinting based methods. To estimate the location of the target, the former one utilizes the propagation model of wireless signals in forms of the received signal strength (RSS), the time of flight (TOF) and/or angle of arrival (AOA)~\cite{HuiLiu2007,Chen2017}.  In contrast, the latter considers the physically measurable properties of WiFi as fingerprints or signatures for each discrete spatial point to discriminate between locations. Due to the wide fluctuation of WiFi signals~\cite{Xue2019} in an indoor environment, the exact propagation model is difficult to obtain, which makes the fingerprinting approach more favorable.   

In fingerprint methods, the received signal strength indicator (RSSI) is widely used as a feature in localization because RSSI can be obtained easily from most WiFi receivers such as mobile phones, tablets, laptops, etc.~\cite{Chen2016,Haochen2017}. However, RSSI has two drawbacks: instability due to fading and multipath effects and device heterogeneity due to the fact that different devices have different RSSIs even at the same position~\cite{Haochen2017}. In order to mitigate these problems, channel state information (CSI) is adopted to provide richer information from multiple antennas and multiple subcarriers~\cite{Haochen2017, Wang2017b}. Although CSI is a more detailed fingerprint to improve the localization accuracy, it is only available with the specific wireless network interface cards (NIC), e.g., Intel WiFi Link 5300 MIMO NIC, Atheros AR9390 or Atheros AR9580 chipset~\cite{Wang2017b}. Therefore, RSSI is still a popular choice in practical scenarios.    

Among WiFi RSSI fingerprinting indoor localization approaches, the probabilistic method is based on statistical inference between the target signal measurement and stored fingerprints using Bayes rule~\cite{Youssef2005}. The RSSI probability density function (PDF) is assumed to have empirical parametric distributions (e.g., Gaussian, double-peak Gaussian, lognormal~\cite{Xu2016}), which may not be necessarily accurate in practical situations. In order to achieve better performance, non-parametric methods~\cite{Kushki2007, C.Figuera2009} did not make no assumption on the RSSI PDF but require a large amount of data at each reference point (RP) to form the smooth and accurate PDF. Beside the probabilistic approach, the deterministic methods use a similarity metric to differentiate the measured signal and the fingerprint data in the dataset to locate the user's position~\cite{He2016}. The simplest deterministic approach is the K nearest neighbors (KNN)~\cite{Bahl2000, YaqinXie2016, DongLi2016} model which determines the user location by calculating and ranking the fingerprint distance measured at the unknown point and the reference locations in the database. Moreover, support vector machine (SVM)~\cite{Brunato2005} provides a direct mapping from RSSI values collected at the mobile devices to the estimated locations through nonlinear regression by supervised classification technique~\cite{C.Gentile2013}. Despite their low complexity, the accuracy of these methods are unstable due to the wide fluctuation of WiFi RSSI~\cite{YaqinXie2016,DongLi2016,Brunato2005}. In contrast to these algorithms, artificial neural network (ANN)~\cite{Haochen2017, Fang2008} estimates location nonlinearly from the input by a chosen activation function and adjustable weightings. In indoor environments, because the transformation  between the RSSI values and the user's locations is nonlinear, it is difficult to formulate a closed form solution~\cite{C.Gentile2013}. ANN is a suitable and reliable solution for its ability to approximate high dimension and highly nonlinear models~\cite{Haochen2017}. Recently, several ANN localization solutions, such as multilayer perceptron (MLP)~\cite{Battiti2002}, robust extreme learning machine (RELM)~\cite{Lu2016}, multi-layer neural network (MLNN)~\cite{Dai2016}, convolutional neural network (CNN)~\cite{Jiao2017}, etc., have been proposed. 

Although having been extensively investigated in the literature, all of the above algorithms still face challenges such as spatial ambiguity, RSSI instability and RSSI short collecting time per location~\cite{Minh2018}. 
To address these challenges, this paper focuses on recurrent neural network (RNN) to determine the user's location by  exploiting the sequential correlation of RSSI measurements. Since the moving speed of the user in an indoor environment is bounded, the temporal information can be used to distinguish the locations that has similar fingerprints. Note that resolving the ambiguous locations has been a common challenge in indoor localization. Some works in literature also exploit the measurements in previous time steps to locate the current location, including the use of Kalman filter~\cite{Au2013, Guvenc2003, Besada2007, Kushki2006} and soft range limited K-nearest neighbors (SRL-KNN)~\cite{Minh2018}. Among them, Kalman filter estimates the most likely current location based on prior measurements, assuming a Gaussian noise of the RSSI and linear motion of the detecting object. However, in real scenarios, these assumptions are not necessarily valid~\cite{YogitaChapre2013}. In comparison, SRL-KNN does not make the above assumptions but requires that the speed of the targeting object is bounded, e.g., from 0.4~m/s to 2~m/s. If the speed of the target is beyond the limit, the localization accuracy of SRL-KNN will be severely impaired. In contrast, our RNN model is trained from a large number of randomly generated trajectories representing the natural random walking behaviours of humans. Therefore, it does not have the assumptions or constraints mentioned above. 

The main contributions of this paper are summarized as follows.
\begin{itemize}
\item[1] According to our knowledge, there is no existing comprehensive RNN solution for WiFi RSSI fingerprinting  with  detailed  analysis  and  comparisons. Therefore,  we  propose  a  complete study of RNN  architectures including  network  structures  and  parameter analysis  of  several  types  of  RNNs,  such as vanilla RNN, long short term memory (LSTM)~\cite{Hochreiter1997}, gated recurrent unit (GRU)~\cite{Cho2014}, bidirectional RNN (BiRNN), bidirectional LSTM (BiLSTM)~\cite{Schuster1997} and bidirectional GRU (BiGRU)~\cite{Zhao2018}.
\item[2] The  proposed  models  are  tested  in  two  different  datasets including an in-house measurement dataset and the published  dataset UJIIndoorLoc~\cite{Torres-Sospedra2014}. The accuracy is compared not only with the other neural network methods, i.e., MLP~\cite{Battiti2002} and MLNN~\cite{Dai2016}, but also some popular conventional methods, i.e., RADAR~\cite{Bahl2000}, SRL-KNN~\cite{Minh2018}, Kernel method~\cite{Kushki2007} and Kalman filter~\cite{Au2013}.
\item[3] Three challenges of WiFi indoor localization, i.e., spatial ambiguity,  RSSI  instability  and  RSSI  short  collecting time per location are discussed and addressed. Furthermore, the other important factors, including the network training  time  requirement,  user’s  speed  variation,  different testing  time  slots  and  historical  prediction  errors, are discussed and analyzed.
\end{itemize}  
\section{Related Works} \label{sec:related_work}
\begin{table*}[!t]
\centering         
\caption{Comparisons of Indoor Localization Experiments Using Machine Learning Techniques} \label{table:AverageErr} 
\begin{tabular}{l c c c c c c c c} 
\hline           
\textbf{Method} & \textbf{Feature} & \textbf{Access point (AP)} & \textbf{Reference point (RP)} & \textbf{Testing Point} & \textbf{Grid Size}  & \textbf{Accuracy}\\ 
MLP \cite{Battiti2002} & RSSI  &  6 &  207 &  50  & 1.7 m & 2.8 $\pm$ 0.1 m\\
DANN \cite{Fang2008} & RSSI  &  15 &  45 &  46  & 2 m & 2.2 $\pm$ 2.0 m\\
RELM \cite{Lu2016} & RSSI  &  8 &  30 &  10  & 3.5 m & 3.7 $\pm$ 3.4 m\\
MLNN \cite{Dai2016} & RSSI  &  9 &  20 &  20  & 1.5 m & 1.1 $\pm$ 1.2 m\\
ConFi \cite{Haochen2017} & CSI  &  1 &  64 &  10  & 1.5 - 2 m & 1.3 $\pm$ 0.9 m\\
Geomagnetic RNN \cite{Ho2017} & Magnetic Information &  - &  629 &  $5\%$ of RPs  & 0.57 m& 1.1 m\\
\hline         
\end{tabular} 
\end{table*}

\begin{figure}[!t]
\centering
\includegraphics[width=0.48\textwidth]{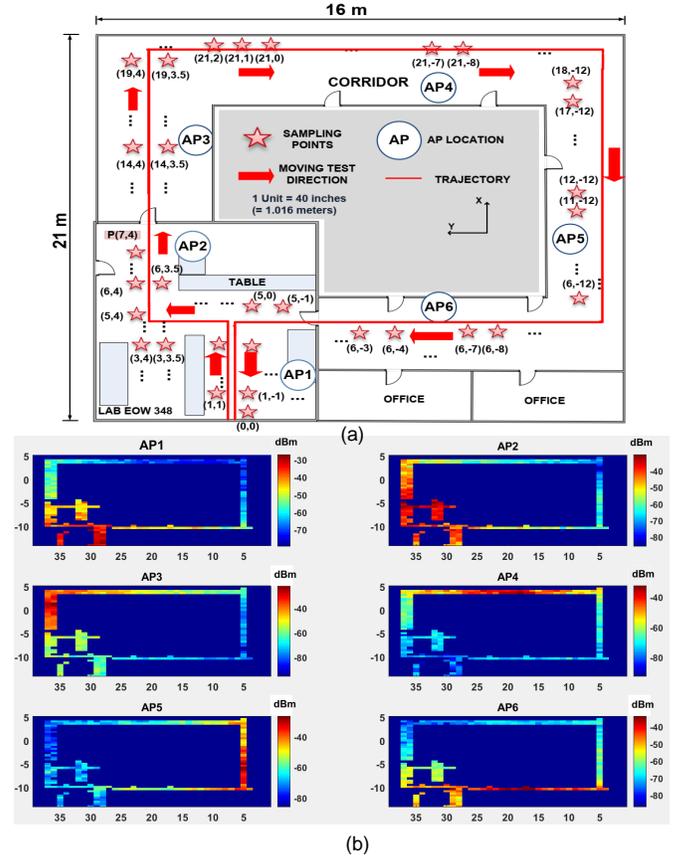}
\caption{(a) Floor map of the test site. The solid red line is the mobile user's walking trajectory with red arrows pointing toward walking direction. (b) Heat map of the RSSI strength from 6 APs used in our localization scheme.}
\label{fig:floor_map}
\end{figure}

The first research on neural network for indoor localization was reported by Battiti \textit{et al.}~\cite{Brunato2005,Battiti2002}. In their work, a multilayer perceptron network with 3 layers consisting one input layer, one hidden layer with 16 neurons and one output layer was implemented to nonlinearly map the output (coordinate) from the input (RSSI). Using 207 RPs for training and 50 random points for validation, the accuracy of this model is $\mathrm{2.82{\pm}0.11~m}$, which is comparable with that of simple KNN algorithm. In order to achieve better performance, multi-layer feed-forward neural network with 3 hidden layers was investigated~\cite{Dai2016}. MLNN is designed with 3 sections: RSSI transformation section, RSSI denoising section and localization section with the boosting method to tune the network parameters for misclassification correction. Other refinements of MLP are proposed in discriminant-adaptive neural network (DANN)~\cite{Fang2008}, which inserts discriminative components (DCs) layer to extract useful information from the inputs. The experiment shows that DANN improves the probability of the localization error below 2.5~m by $\mathrm{17\%}$ over the conventional KNN RADAR~\cite{Bahl2000}. As all of the above methods are time-consuming in the training phase, robust extreme learning machine~\cite{Lu2016} is proposed to increase the training speed of the feedforward neuron networks. RELM consists of a generalized single-hidden layer with random hidden nodes initialization and kernelized formulations with second order cone programming. The experiment illustrates that the training speed of RELM is more than 100 times faster than the conventional machine learning methods~\cite{Huang2012} and the localization accuracy is increased by $\mathrm{40\%}$.    

Although the feedforward neural networks are simple and easy to implement, they cannot extract the useful information efficiently from the noisy WiFi signal, leading to a limited accuracy. Therefore, more complicated neural networks, e.g., CNN and RNN, were adopted for indoor WiFi localization. ConFi~\cite{Haochen2017} proposed a three layers CNN to extract the information from WiFi channel state information. CSI from different subcarriers at different time is arranged into a matrix, which is similar to an image. ConFi is trained using the CSI feature images collected at a number of RPs. The localization result is the weighted centroid of RPs with high output values. The experiment illustrates that ConFi outperforms conventional KNN RADAR~\cite{Bahl2000} by $\mathrm{66.9\%}$ in terms of mean localization error. However, as mentioned above,  CSI is only accessible for some specific wireless network interface cards. Consequently, such algorithms cannot be generically implemented. Recently, RNN is used for indoor localization. Ref.~\cite{Yuan2017} proposed a simple RNN with 1 time step and 2 hidden layers. In their work, RNN classifies 42 RPs based on the RSSI readings from 177 APs. The classification accuracy is $\mathrm{82.47\%}$. A more efficient RNN is published in~\cite{Ho2017}. In that work, the RNN model uses 200 neurons, mean squared error (MSE) loss function, and 20 time step traces as the input for the network. Geomagnetic data is used as fingerprints instead of the wireless data. A million traces of various pedestrian walking patterns are generated with $\mathrm{95\%}$ of them being used for training and $\mathrm{5\%}$ for validation. The achieved localization errors range from 0.441 to 3.874~m with the average error being 1.062~m. In addition to the geomagnetic data, light intensity is also utilized in the deep LSTM network~\cite{Wang2018} to estimate the indoor location of the target mobile device. Their 2-layer LSTM exploits temporal information from bimodal fingerprints, i.e., magnetic field and light intensity data, through recursively mapping the input sequence to the label of output locations. The accuracy is reported as $\mathrm{82\%}$ of the test locations with location errors around 2~m and the maximum error being around 3.7~m.     

Table~\ref{table:AverageErr} summarizes the experimental set-up and the results from the above mentioned neural network methods. Here the number of access points (APs), RPs and testing points vary between those experiments and the grid size is defined as the distance between two consecutive RPs. 

In general, these methods provide acceptable accuracy from 1~m to 3~m but none of them have sufficiently investigated the three problems of using RSSI as fingerprints. Therefore, we propose several RNN solutions, such as vanilla RNN, LSTM, GRU, BiRNN, BiLSTM and BiGRU, to solve the three RSSI fingerprinting challenges. Our localization results are compared not only with the other neural network methods, i.e., MLP~\cite{Battiti2002} and MLNN~\cite{Dai2016}, but also some conventional methods, i.e., RADAR~\cite{Bahl2000}, SRL-KNN~\cite{Minh2018}, Kernel method~\cite{Kushki2007} and Kalman filter~\cite{Au2013}. 
\section{RNN Methods} \label{sec:system}

\subsection{Recurrent Neural Network Overview}
A recurrent neural network is a class of artificial neural network, where the output results depend not only on the current input value but also on the historical data~\cite{Ho2017}. RNN is often used in situations, where data has a sequential correlation. In the case of indoor localization, the current location of the user is correlated to its previous locations as the user can only move along a continuous trajectory. Therefore, RNN exploits the sequential RSSI measurements and the trajectory information to enhance the accuracy of the localization. Several RNN models have been proposed. The vanilla RNN~\cite{Zachary2015}, the simplest RNN model, has limited applications due to vanishing gradient during the training phase~\cite{Chung2014}. To mitigate this effect, long short term memory~\cite{Hochreiter1997} creates an internal memory state which adds the forget gate to control the time dependence and effects of the previous inputs. GRU~\cite{Cho2014} is similar to LSTM but only consists of the update and reset gate instead of the forget, update and output gate in LSTM. BiRNN~\cite{Zachary2015}, BiLSTM~\cite{Schuster1997} and BiGRU~\cite{Zhao2018} are the extensions of the traditional RNN, LSTM and GRU respectively, which not only utilize all available input information from the past but also from the future of a specific time frame. In the following section we will describe the details about the proposed WiFi RSSI indoor localization system using RNN models. 
\begin{figure*}[!t]
\centering
\includegraphics[width=\textwidth]{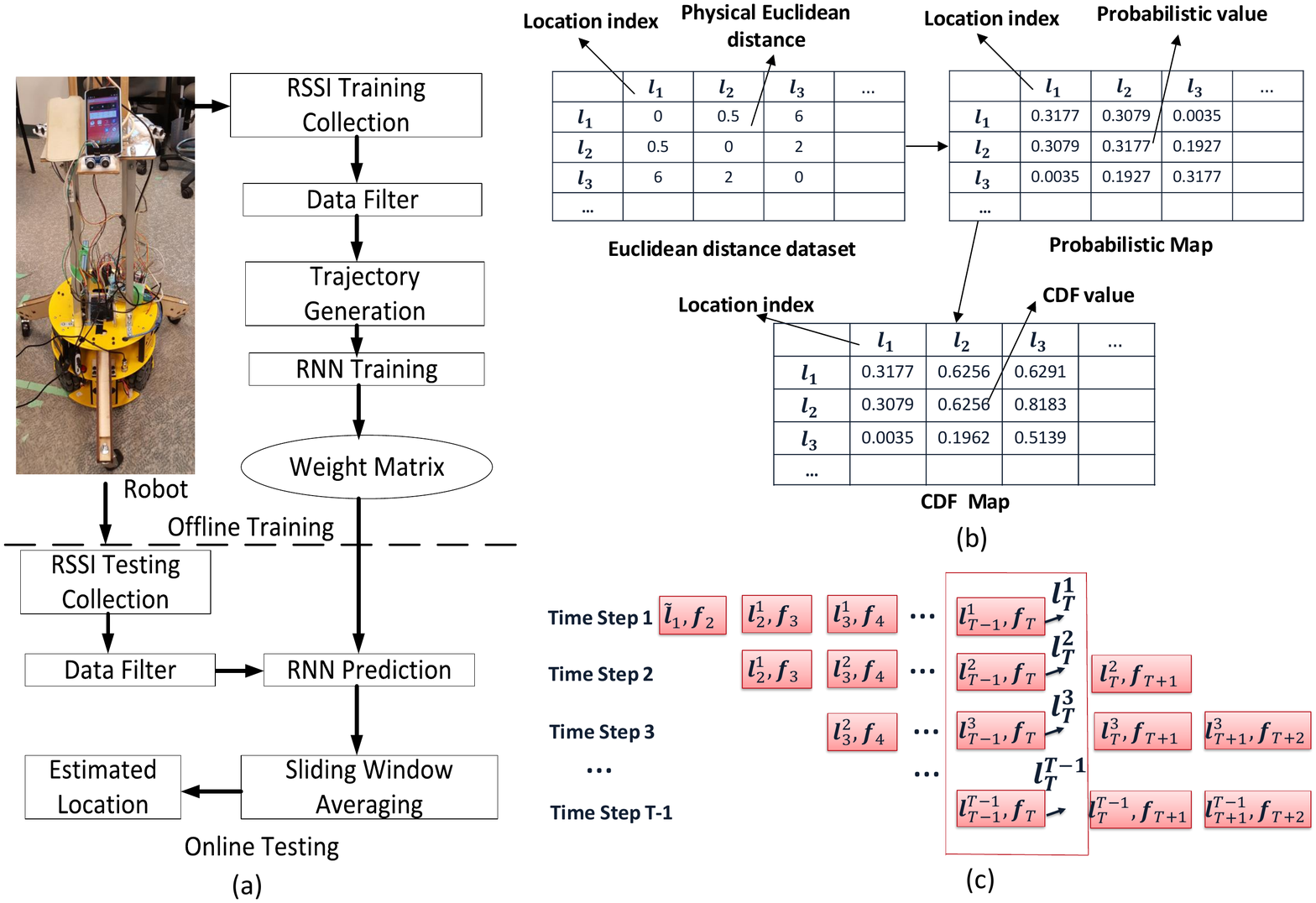}
\caption{(a) Localization process of the proposed RNN system. (b) Trajectory generation process. (c) Sliding window averaging in online testing phase.}
\label{fig:map_gen}
\end{figure*}

\subsection{Proposed Localization System}
The development of our localization RNN models is divided into two phases: a training phase (offline phase) and a testing phase (online phase). In the training phase, RSSI readings at each predefined RP  are collected into a database. Assuming that the area of interest has $P$ APs and $M$ RPs, we consider each RP $i$ at its physical location $\bm{l}_{i}(x_{i},y_{i})$ having a corresponding fingerprint vector  $\bm{f}_{i}=\{F^{i}_{1}, F^{i}_{2},...,F^{i}_{N}\}$, where $N$ is the number of available features (RSSIs) from all APs at all carrier frequencies and $F^{i}_{j}(1 \leq j \leq N)$ is the $j$-th feature at RP $i$.  During the training phase, a number of scans are collected at a single RP while in the testing phase fewer scans of RSSIs readings is collected as the user is mobile in practical scenario. Fig.~\ref{fig:floor_map}(a) illustrates the localization map with 6~APs, 365~RPs and 175~testing locations, while Fig.~\ref{fig:floor_map}(b) shows the heat map of these 6~APs. 
\begin{figure*}[!t]
\centering
\includegraphics[width=\textwidth]{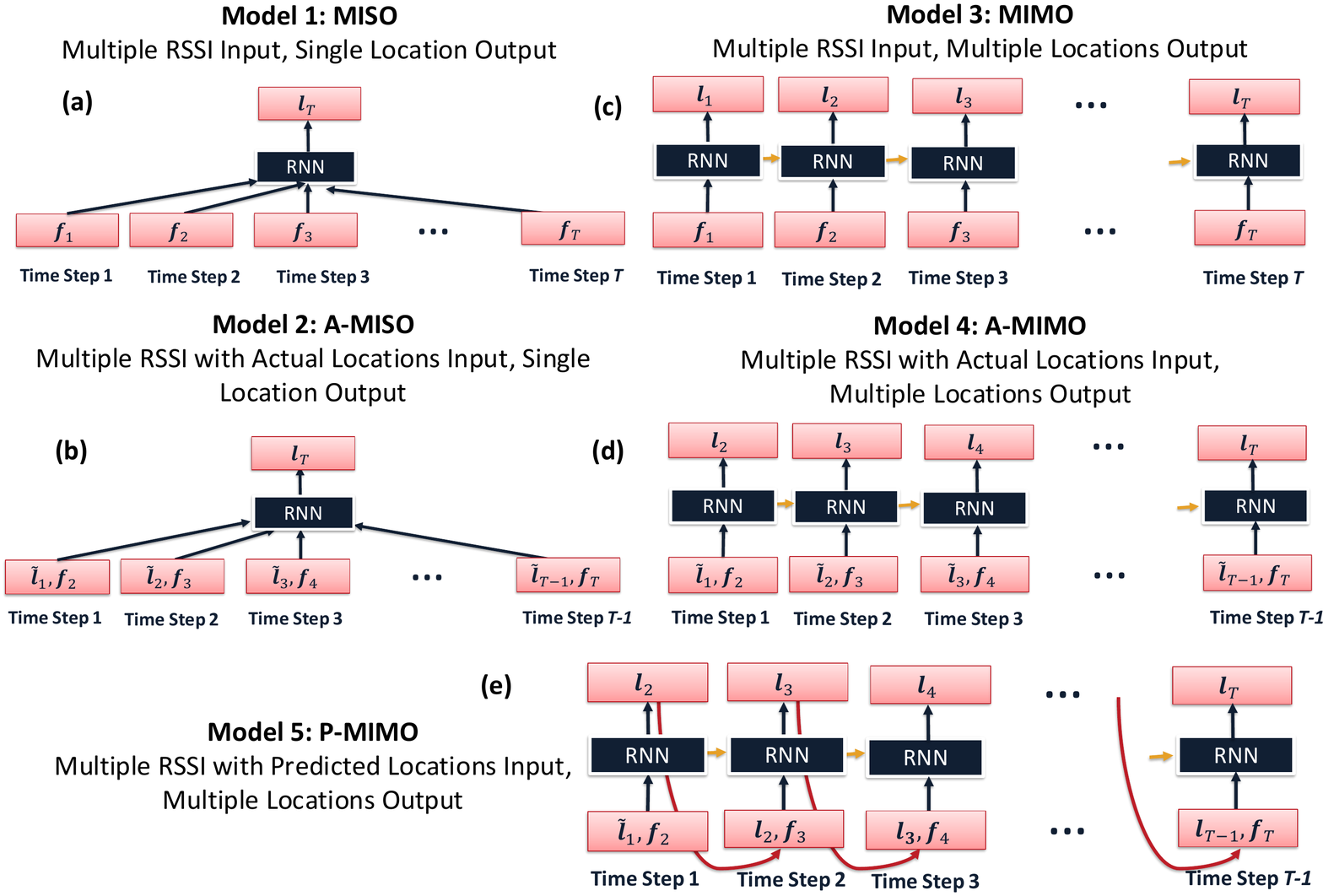}
\caption{Proposed RNN models.}
\label{fig:Diff_Struct}
\end{figure*}
The architecture of the proposed RNN system is presented in Fig.~\ref{fig:map_gen}(a). Details of the process are described as follows. 

\subsubsection{Data Filter}  \label{sub:data_filter} 
\begin{figure}[!t]
\centering
\includegraphics[width=0.5\textwidth]{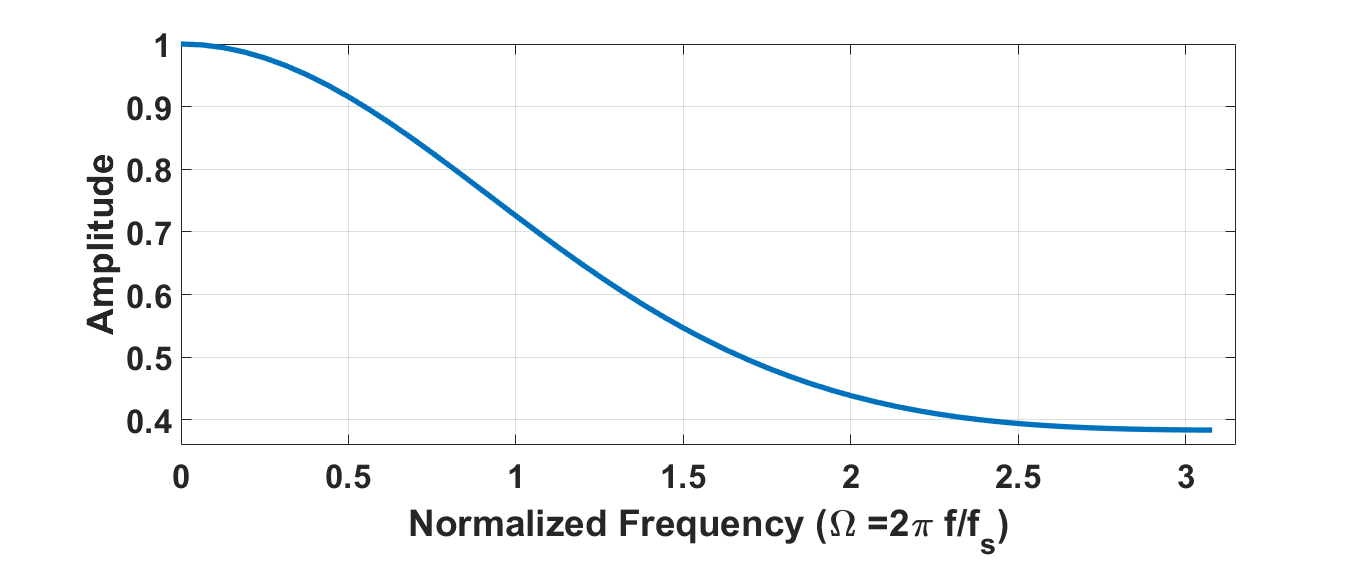}
\caption{Weighted average filter transfer function.}
\label{fig:Transfer}
\end{figure}

During the training phase, RSSIs at RPs are collected by a mobile device mounted on an autonomous driving robot and stored in a database. As expected, the RSSI measurements often experience substantial fluctuations due to dynamically changing environments such as human blocking and movements, interference from other equipment and devices, receiver antenna orientation, etc.,~\cite{YogitaChapre2013}. For example, the mobile device was placed at $P(7,4)$ as shown in Fig.~\ref{fig:floor_map}(a). The experiment was conducted during working hours when many students used WiFi and walked around the lab. The standard deviation of maximum RSSI over 100 consecutive RSSI readings was 5.5~dB, and $\mathrm{5\%}$ of the measurements (5 readings) could not be detected. In those cases, the mobile device missed the beacon frame packets sent by the router. In order to filter out those outliers, we adopt the iterative recursive weighted average filter \cite{Pei2015} with 3 taps and 5 different weighted factors, i.e., $\beta_{1} = 0.8$, $\beta_{2} = 0.2$, $\beta_{3} = 0.8$, $\beta_{4} = 0.15$, $\beta_{5} = 0.05$. This weighted filter has the form of a low pass filter as shown in Fig.~\ref{fig:Transfer}. In both the training and testing phase, the filter is used in the same way with fixed taps and weighted factors. The effectiveness of the filter will be studied in Section~\ref{sec:sim_result}.  

\subsubsection{Trajectory Generation} \label{sub:map_gen} 
The RSSIs at the output of the data filter will be used to generate random training trajectories under the constraints that the distance between consecutive locations is bounded by the maximum distance a user can travel within the sample interval in practical scenarios. Fig.~\ref{fig:map_gen}(b) illustrates the trajectory generation process. Firstly, the physical Euclidean distance between each location and the rest of the database is calculated to form a Euclidean distance dataset. Secondly, based on that, the probabilistic map will be generated to represent the probability of a location ($P(\bm{l}_i)$) that will become the next location of the user in the trajectory. Since the moving speed of an indoor user is limited, the locations which are near to the previous locations should have higher probability to become the next location in the trajectory than the further locations. The user location will be updated in every consecutive sampling time interval $\Delta{t}$. Therefore, the maximum distance which the user can move in $\Delta{t}$ is $\sigma = v_{max} \times \Delta{t}$, where the maximum speed $v_{max}$ is chosen to be larger than the human indoor normal speed (from 0.4~m/s to 2~m/s~\cite{Browning2006,Email2007}). The normalized probability $P(\bm{l}_i)$ is calculated as follows.
\begin{equation} \label{eq:p_l}
P(\bm{l}_i) = \frac{1}{2\sigma^{2}(1-e^{\frac{d_{max}^2}{2\sigma^2}})}\exp(-\frac{(x_{i}-x_{pre})^2+(y_{i}-y_{pre})^2}{2\sigma^2})
\end{equation} 
where $(x_{pre},y_{pre})$ is the most recent location of $\bm{l}_i$, $d_{max}$ is the maximum distance between the considered location $\bm{l}_i$ and the furthest location in the interested area. Eq.~\eqref{eq:p_l} has the form of a Gaussian distribution with the mean being the previous location and the standard deviation being $\sigma$. 
All of the locations having the same physical distance with $\bm{l}_i$ will get the same probability to be chosen as the next point on the trajectory. From the probabilistic map, a cumulative distribution function (CDF) map is built by summing the $P(\bm{l}_i)$ for each location $\bm{l}_i$. Finally, in order to get the next location $\bm{l}_j$ of a any location $\bm{l}_i$ in the trajectory, a random number $R$ ($0<R<1$) is picked. $\bm{l}_j$ is the location which has the value in the CDF map being closest with $R$.

\subsubsection{Proposed RNN Models} \label{sub:rnn_model} 
The proposed RNN architecture is trained by the data from consecutive locations in a trajectory to exploit the time correlation between them. Each location in a trajectory appears at a different time step. The length of a trajectory, or equivalently the number of time steps, defines the memory length $T$ as illustrated in Fig.~\ref{fig:Diff_Struct}. The number of time steps $T$ significantly impacts the performance of RNN because all of the weights and hidden states will be saved at every time step during a training trajectory~\cite{Zachary2015}. A larger $T$ value will incorporate more information from the past but will accumulate more localization errors. The optimal $T$ value will be chosen by the experiment in Section~\ref{sec:sim_result}.   

Fig.~\ref{fig:Diff_Struct} illustrates 5 different proposed RNN models labeled from model 1 to model 5. Model 1 has a multiple input single output  (MISO) structure where several RSSI readings  ($\bm{f}_i$) from the previous time steps are fed into the network to get the single location at time step $T$ ($\bm{l}_T$). Model 2, similar to model 1, has a multiple input single output (A-MISO) structure. However, model 2 takes in actual previous step locations as well as RSSIs unlike model 1. The actual locations are the ground truth locations $\tilde{\bm{l}}$.

By comparison, model 3 to model 5 are multiple input multiple output (MIMO) structures where several RSSI readings from multiple time steps are fed to the network to get multiple output locations $\bm{l}$. Model 3 has a MIMO structure, where multiple RSSI inputs $\bm{f}_i$ produce multiple output locations $\bm{l}_i$. In contrast,  model 4 takes in multiple RSSIs with multiple actual locations for the input and produces multiple locations for the output, denoted as A-MIMO.

Model 5 takes in multiple RSSIs and multiple previous predicted locations for the input and produces multiple locations for the output, denoted as P-MIMO. Specifically, in model 4: A-MIMO, the location used in the training is the ground truth from the dataset. In model 5: P-MIMO, the location value is the predicted location ($\bm{l}_i$) from the previous time step. Note that in the testing phase, both model 2 and model 4 use the predicted locations from the previous time steps as input because ground truth is not available. 

The objective of RNN training is to minimize the loss function $\mathcal{L}(\bm{l},\tilde{\bm{l}})$ defined as the Euclidean distance between the output $\bm{l}$ and the target $\tilde{\bm{l}}$ using the backpropagation algorithm~\cite{Zachary2015}. In single output models such as MISO and A-MISO,  the loss function is given by
\begin{equation} \label{eq:loss1}
 \mathcal{L}(\bm{l},\tilde{\bm{l}}) = ||\bm{l}_T - \tilde{\bm{l}}_T||_2.
\end{equation}  
In contrast, the multiple output models MIMO, A-MIMO and P-MIMO adopt a loss function expressed by
\begin{equation} \label{eq:loss2}
 \mathcal{L}(\bm{l},\tilde{\bm{l}}) = \frac{\sum_{i=1}^{T} ||\bm{l}_i - \tilde{\bm{l}}_i||_2}{T}.
\end{equation}    

\subsubsection{Sliding Window Averaging}
MIMO, A-MIMO and P-MIMO have the output locations appearing in every time step (Subsection~\ref{sub:rnn_model}). In the online testing phase, the output location $\bm{l}_T$ will appear in several time steps as shown in Fig.~\ref{fig:map_gen}(c). $\bm{l}_i^j$ is the output location $\bm{l}_i$ of time step $j$. At the output time step $T-1$, we have a set of output location $(\bm{l}_T^1, \bm{l}_T^2, ..., \bm{l}_T^{T-1})$ from $T-1$ previous steps. In each output time step, the accuracy of the targeted output result $\bm{l}_T$ can be slightly different due to the difference in length of previous historical information. For example, in Fig.~\ref{fig:map_gen}(c), at output time step 1, $\bm{l}^1_T$ is estimated with the information of $T-1$ previous steps. However, at output time step 2, the number of previous steps for $\bm{l}^2_T$ decrease to $T-2$; and at time step $T-1$, $\bm{l}^{T-1}_T$ is predicted with no previous step. Therefore, the sliding window averaging can average the error of the predicted location $\bm{l}_T$ in multiple output steps and increase the localization accuracy. The final output result $\bm{l}_T$ will be the average of the above output set:
\begin{equation} \label{eq:average}
\bm{l}_T = \frac{\sum_{j=1}^{T-1} \bm{l}_T^j}{T-1}.
\end{equation}
 
\section{Database And Experiments} \label{sec:database}
All experiments have been carried out on the third floor of Engineering Office Wing (EOW), University of Victoria, BC, Canada. The dimension of the area is 21~m by 16~m. It has three long corridors as shown in Fig.~\ref{fig:floor_map}(a). There are 6 APs and 5 of them provide 2 distinct MAC address for 2.4~GHz and 5~GHz communications channels respectively, except for one that only operates on 2.4~GHz frequency. Equivalently, in every scan, 11 RSSI readings from those 6 APs can be collected. 

The RSSI data for both training and testing will be collected using an autonomous driving robot. The 3-wheel robot as shown in Fig.~\ref{fig:map_gen}(a) has multiple sensors including a wheel odometer, an inertial measurement unit (IMU), a LIDAR, sonar sensors and a color and depth (RGB-D) camera. It can navigate to a target location within an accuracy of  $\mathrm{0.07{\pm}0.02~m}$. The robot also carries a mobile device (Google Nexus 4 running Android 4.4) to collect WiFi fingerprints. The dataset for offline training was collected by the phone-carrying robot at 365~RPs. At each location, 100 scans of RSSI measurements ($S_{1}=100$) were collected. To build the dataset, at each location in the training trajectory, we randomly choose one out of 100 stored RSSIs as the RSSI associated with the location. There are total 365,000 random generated training trajectories following the proposed method in Subsection~\ref{sub:map_gen}. This approximates well the user's random walk property and helps to reduce the spatial ambiguity.  The initial position of the user in the whole testing trajectory is known.

In the online phase, the robot moved along a pre-defined route (Fig.~\ref{fig:floor_map}(a)) with an average speed around 0.6~m/s. The robot will collect RSSI at 175 testing locations along the trajectory. At each location, only 1 or 2 RSSI scans ($S_{2}=1$ or $2$) will be collected and transmitted to a server in real time. The server will predict the user's position according to the proposed algorithms and the prediction accuracy will be calculated.

\section{Results And Discussions} \label{sec:sim_result}
The initial setup for the proposed RNN system follows the parameters presented in Table~\ref{table:setup}. All of the results below are presented after 10-fold tests with a total of 365,000 random training trajectories.      
\begin{table}
\centering         
\caption{Initial setup parameters for RNN system} \label{table:setup} 
\begin{tabular}{c c}          
\hline
\textbf{Category} & \textbf{Value}  \\ 
K-fold tests & 10 \\
RNN type & LSTM \\
Memory length ($T$) & 10 \\
Model & P-MIMO\\
Loss function & RMSE\\ 
Hidden layer (HL)  & 2 \\
Number of neurons for each HL & 100\\
Dropout & 0.2 \\
Optimizer & Adam \\
Learning rate & 0.001 \\
Number of training trajectory for 1-fold test & 10,000 \\
Number of training epochs & 1000 \\
$\sigma$ & 2 m \\
$\Delta_t$ & 1 s \\
$d_{max}$ & 2 m \\
 \hline  
\end{tabular} 
\end{table}

\begin{table*}[!t]
\centering         
\caption{Average localization errors} \label{table:Error1} 
\begin{tabular}{l c c c c c c c} 
\hline           
\textbf{Method} & \textbf{MISO LSTM} & \textbf{A-MISO LSTM} & \textbf{MIMO LSTM} & \textbf{A-MIMO LSTM} & \textbf{P-MIMO LSTM}\\ 
Average Error (m) &1.05 $\pm$ 0.78 &  0.92 $\pm$ 0.75 &  0.80 $\pm$ 0.67 &  0.91 $\pm$ 0.75 & 0.75 $\pm$ 0.64\\
\textbf{Method} & \textbf{RADAR\cite{Bahl2000}} &  \textbf{MLNN \cite{Dai2016}}  &  \textbf{MLP \cite{Battiti2002}} & \textbf{Kernel Method\cite{Kushki2007}} & \textbf{Kalman  Filter\cite{Au2013}}\\ 
Average Error (m) & 1.13 $\pm$ 0.86  &  1.65 $\pm$ 1.20 & 1.72 $\pm$ 1.17 & 1.10 $\pm$ 0.84 & 1.47  $\pm$ 1.2 \\
\hline         
\end{tabular} 
\end{table*}
\subsection{Filter Comparison}
\begin{figure}[!t]
\centering
\includegraphics[width=0.52\textwidth]{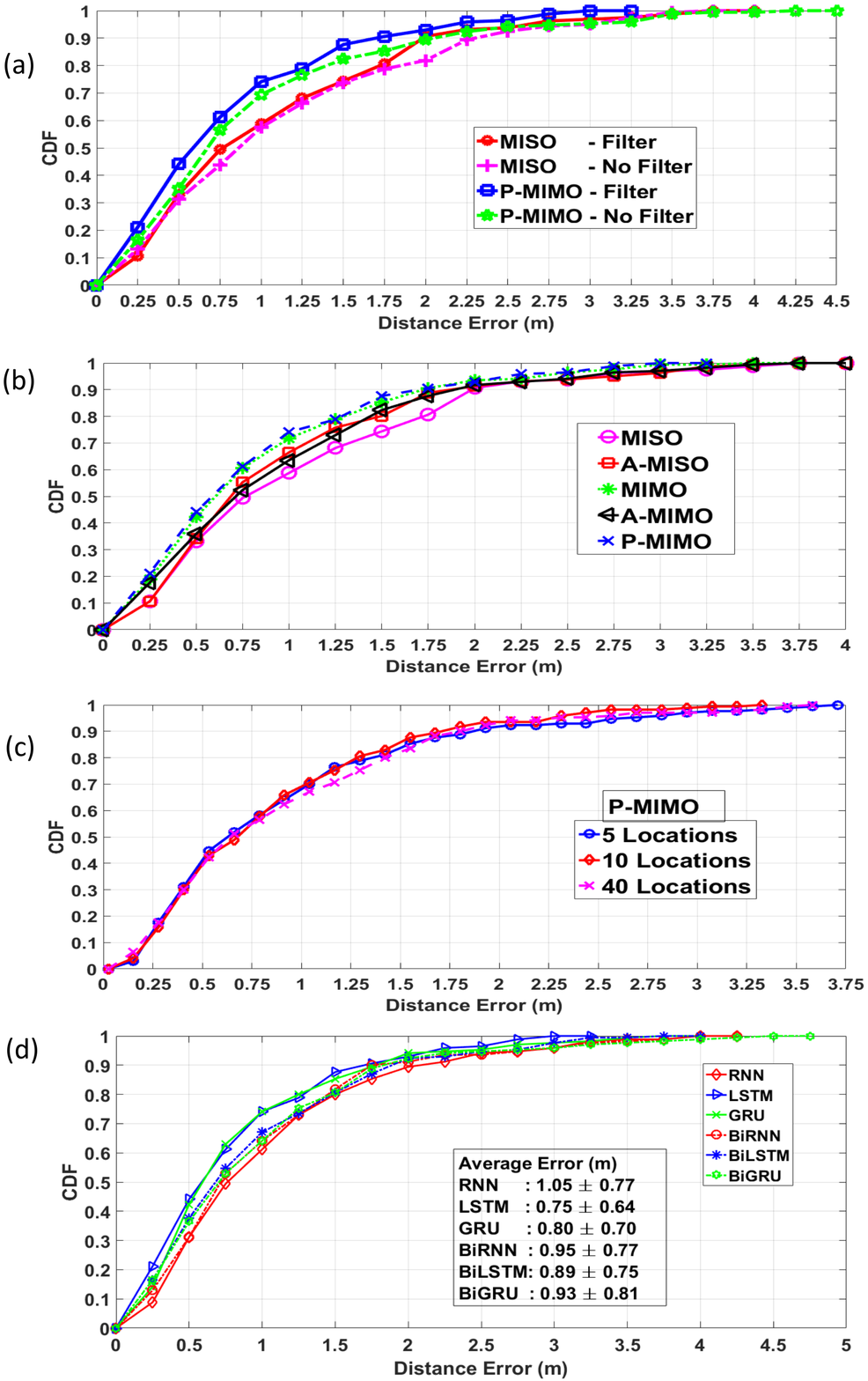}
\caption{The CDF of the localization error of (a) Filter and no filter cases (b) 5 different RNN models (c) Different memory lengths in RNN structure (d) RNN, LSTM, GRU, BiRNN, BiLSTM and BiGRU with P-MIMO model.}
\label{fig:CDF_compare1}
\end{figure}   
Fig.~\ref{fig:CDF_compare1}(a) compares the CDF of localization errors among the proposed RNN models (i.e., MISO LSTM and P-MIMO LSTM) with and without weighted average filter (Subsection~\ref{sub:data_filter}) in both offline training and online testing phases. In P-MIMO, the filter decreases the maximum error from 4.5~m to 3.25~m. In addition, $\mathrm{80\%}$ of the error is within 1.25~m with the filter while without filter the value increases to 1.5~m. In MISO, the filter also leads to better performance with $\mathrm{80\%}$ of the error being within 1.75~m compared with 2~m without filter. In A-MISO, MIMO and A-MIMO, the weighted average filter also provides consistently good results. The localization accuracy of those methods has approximately $15\%$ improvement with filter.
Because of the apparent improvements, we adopt the filter to all the models proposed in this paper.

\subsection{Model Comparison}
Table~\ref{table:Error1} illustrates the average errors of all proposed models in Subsection~\ref{sub:rnn_model}. Among them, P-MIMO achieves the best performance with an accuracy of $\mathrm{0.75{\pm}0.64~m}$.  MIMO is the second best performer with the average error of $\mathrm{0.80{\pm}0.67~m}$. MISO has the worst accuracy with the error of $\mathrm{1.05{\pm}0.78~m}$. Fig.~\ref{fig:CDF_compare1}(a) compares the CDF errors among these five models. MIMO and P-MIMO  consistently show the dominating accuracy with $\mathrm{80\%}$ of the errors within 1.2~m, compared with 1.7~m of MISO. Furthermore, the maximum error of P-MIMO is 3.25~m which is lower than the 4~m obtained from MISO. In the rest of the paper, P-MIMO is chosen for further performance study.    

\subsection{Hyper-parameter Analysis}
This subsection provides a detailed study of choosing the optimal hyper-parameters for the proposed model P-MIMO. Although the 5 proposed models, i.e., MISO, A-MISO, MIMO, A-MIMO, P-MIMO, have different input and output, their general neural network structures are mostly similar. Therefore, the procedure to select the optimal hyper-parameters of P-MIMO will also be valid for the other proposed models including MISO, A-MISO, MIMO and A-MIMO.
\subsubsection{Memory Length $T$} \label{sub:mem_exp}
\begin{figure}[!t]
\centering
\includegraphics[width=0.5\textwidth]{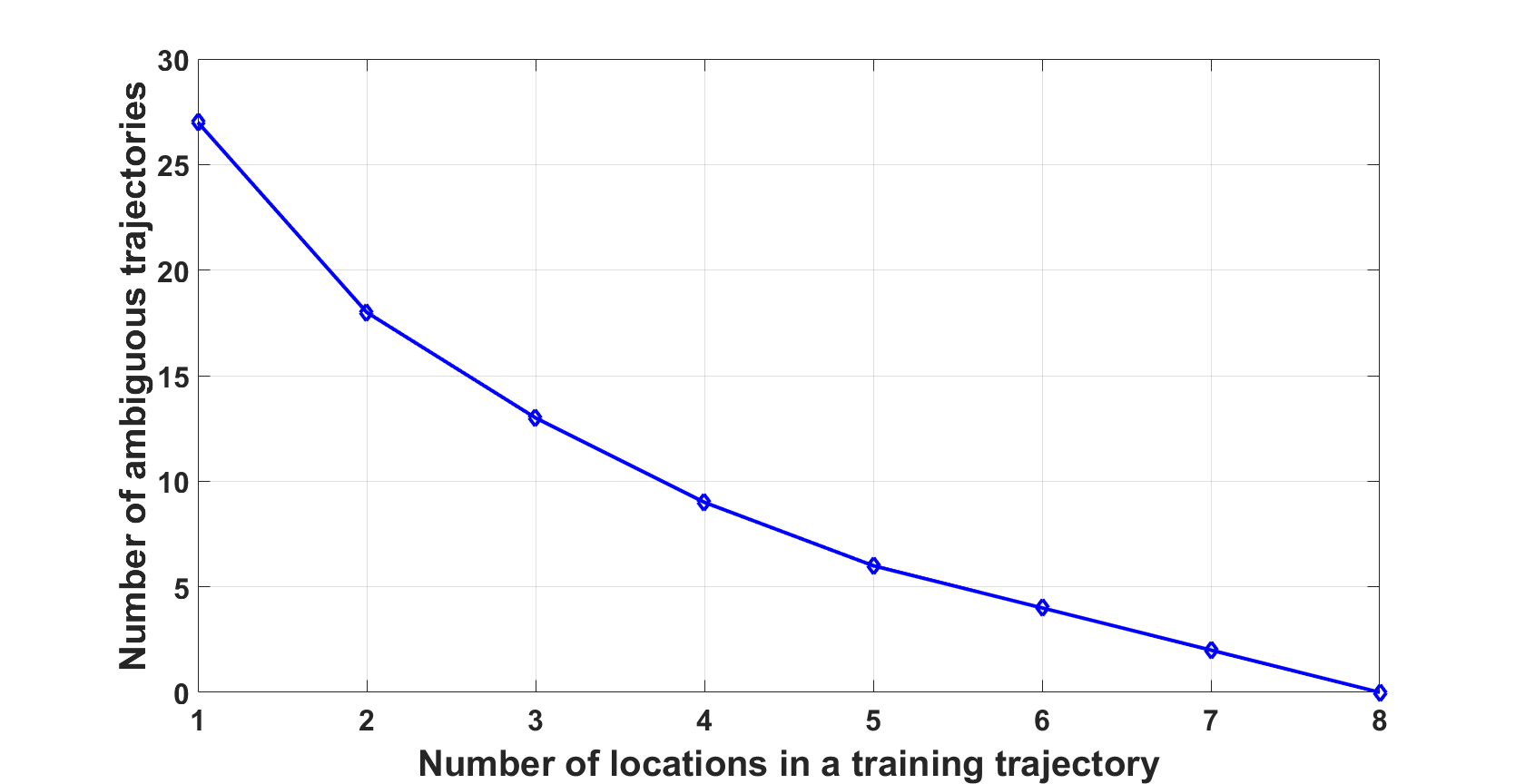}
\caption{Average number of ambiguous trajectories with different number of locations in a training trajectory}
\label{fig:Am_Location}
\end{figure}
Fig.~\ref{fig:CDF_compare1}(c) shows the results of P-MIMO with different memory lengths (Subsection~\ref{sub:rnn_model}), i.e., a training trajectory has 5, 10 or 40 locations. The performance of all three cases is comparable. The 10 time steps training trajectory has a slightly better accuracy with the maximum error being only 2.9~m, compared with 3.5~m and 3.75~m in the case of 40 time steps and 5 time steps respectively.         

The theoretical explanation is as follows. A location $\bm{l}_{j}$ is defined as an ambiguous point of $\bm{l}_{i}$ if their physical distance is larger than the grid size but their two vectors $\bm{f}_{i}$ and $\bm{f}_{j}$ have high Pearson correlation coefficient above the correlation threshold. Besides, two locations are defined as physical neighbours if the physical distance between them is less than or equal to the grid size. The correlation threshold is chosen based on the average correlation coefficients between $\bm{l}_{i}$ and all of its physical nearest neighbours, i.e., approximately 0.9 in our database. Then all non-nearest-neighbour locations whose correlation coefficient above this threshold are considered as ambiguous points. Pearson correlation coefficient $\rho(\bm{f}_{i},\bm{f}_{j})$  between $\bm{f}_{i}$  and $\bm{f}_{j}$ can be calculated as follows  
\begin{equation} \label{eq:pearson}
\rho(\bm{f}_{i},\bm{f}_{j}) = \frac{1}{N-1} \sum_{n=1}^{N} (\frac{F^{i}_{n}-\mu_i}{\delta_i}) (\frac{F^{j}_{n}-\mu_j}{\delta_j}) 
\end{equation}
where $\mu_i$, $\mu_j$ are the mean of $\bm{f}_{i}$ and $\bm{f}_{j}$ respectively, $\delta_i$, $\delta_j$ are the standard deviation of $\bm{f}_{i}$ and $\bm{f}_{j}$ respectively. Similar to the definition of the ambiguous location, 2 trajectories are defined as ambiguous if they include different locations but the combinations of their fingerprints have a high Pearson correlation coefficient. Fig.~\ref{fig:Am_Location} demonstrates the advantage of the proposed LSTM model which exploits the sequential trajectory locations compared with the conventional single point prediction. In the case of single point prediction ($1$ location in a training trajectory), the average number of ambiguous locations in our database are 27. If we increase the number of locations in a trajectory for training, the number of ambiguous trajectories decreases significantly. If a training trajectory has more than 8 locations, there will be no ambiguity in our database. Therefore, the memory length configured as 10 locations is reasonable to remove all the ambiguity. 
\subsubsection{Number of Hidden Layers and Neurons}
\begin{figure}[!t]
\centering
\includegraphics[width=0.52\textwidth]{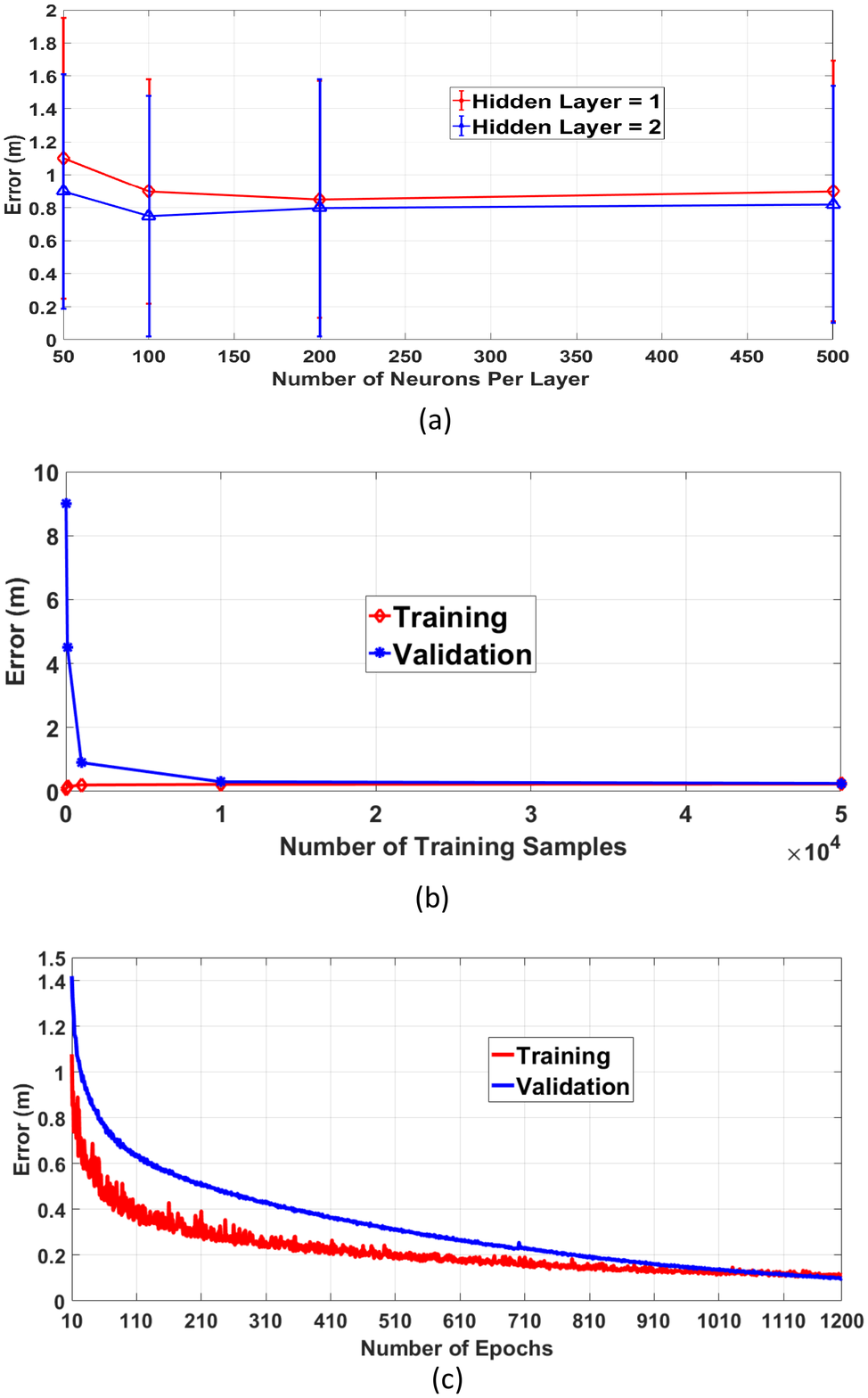}
\caption{(a) Average localization errors of P-MIMO LSTM with different number of hidden layers and neurons per layer. (b) Learning curve of P-MIMO LSTM with the average localization error vs. the number of training trajectory samples. (c) Learning curve of P-MIMO LSTM with the average localization error (the training trajectory samples = $\mathrm{10^4}$) vs. the number of running epochs.}
\label{fig:hyper}
\end{figure}

Fig.~\ref{fig:hyper}(a) illustrates the average localization errors of P-MIMO with different number of hidden layers and neurons per layer. In general, adopting 2 hidden layers leads to better accuracy than 1 hidden layer. In the 1 hidden layer model, using 200 neurons results the best accuracy of $\mathrm{0.85{\pm}0.72~m}$. Increasing more neurons does not result in better performance. In comparison, the best accuracy of 2 hidden layers model is $\mathrm{0.75{\pm}0.64~m}$ when the number of neurons per layer is 100. In summary, 2 hidden layers and 100 neurons per layer are the optimal parameters for our proposed model.  
\subsubsection{Number of Training Trajectories and Training Epochs} \label{sub:training_require}
The learning curve can determine the minimum required number of training trajectory samples and training epochs for the proposed RNN models. Fig.~\ref{fig:hyper}(b) shows the relationship between the training and validation errors vs. the number of training trajectory samples of the proposed P-MIMO model. According to Fig.~\ref{fig:hyper}(b), overfitting will be mitigated if the number of training samples increases to $\mathrm{10^4}$. Therefore, $\mathrm{10^4}$ is the minimum number for the random training trajectories to feed to P-MIMO in the training phase. Fig.~\ref{fig:hyper}(c) illustrates the relationship between the training and cross-validation errors vs. the number of running epochs of P-MIMO LSTM when the training trajectory samples is $\mathrm{10^4}$. When the cross-validation error is at a minimum on the graph, the minimum number of epochs is approximately 1,000. 
\subsubsection{Learning Rate, Optimization Algorithm and Dropout Rate}
\begin{table}[!t]
\centering         
\caption{Different learning rates and optimization algorithms} \label{table:opt_alg} 
\begin{tabular}{c c c}          
\hline
\textbf{Optimization algorithm} & \textbf{Learning rate}  & \textbf{Average Error (m)}\\ 
Adam & 0.01 & 1.0 $\pm$ 0.78\\
Adam & 0.001 & 0.75 $\pm$ 0.64\\
Adam & 0.0001 & 0.80 $\pm$ 0.62\\
SGD & 0.01 & 1.52 $\pm$ 1.32\\
SGD & 0.001 & 1.92 $\pm$ 1.52\\
SGD & 0.0001 & 1.85 $\pm$ 1.25\\
RMSProp & 0.01 & 1.05 $\pm$ 0.95\\
RMSProp & 0.001 & 0.88 $\pm$ 0.72\\
RMSProp & 0.0001 & 0.85 $\pm$ 0.68\\
 \hline  
\end{tabular} 
\end{table}

\begin{table}[!t]
\centering         
\caption{Different dropout rates} \label{table:dropout} 
\begin{tabular}{c c}          
\hline
\textbf{Dropout rate} & \textbf{Average Error (m)}\\ 
0.1 & 0.72 $\pm$ 0.68\\
0.2 & 0.75 $\pm$ 0.64\\
0.3 & 0.87 $\pm$ 0.74\\
0.4 & 0.92 $\pm$ 0.73\\
 \hline  
\end{tabular} 
\end{table}

Table~\ref{table:opt_alg} shows the localization errors when the learning rates and optimization algorithms are varied.  Clearly, optimizer ADAM with the learning rate 0.001 provides the best results with the average error of $\mathrm{0.75{\pm}0.64~m}$. Table~\ref{table:dropout} demonstrates the accuracy of different dropout rates. With the dropout rate equals or less than 0.2, the performance of P-MIMO is mostly unchanged, i.e., $\mathrm{0.72{\pm}0.68~m}$ at a dropout rate of 0.1 and $\mathrm{0.75{\pm}0.64~m}$ at a dropout rate of 0.2. After the dropout rate increases to above 0.2, the accuracy is deteriorated and reaches the bottom of $\mathrm{0.92{\pm}0.73~m}$ at a dropout rate of 0.4.     
\subsection{RNNs Comparison}
\begin{figure}[!t]
\centering
\includegraphics[width=0.52\textwidth]{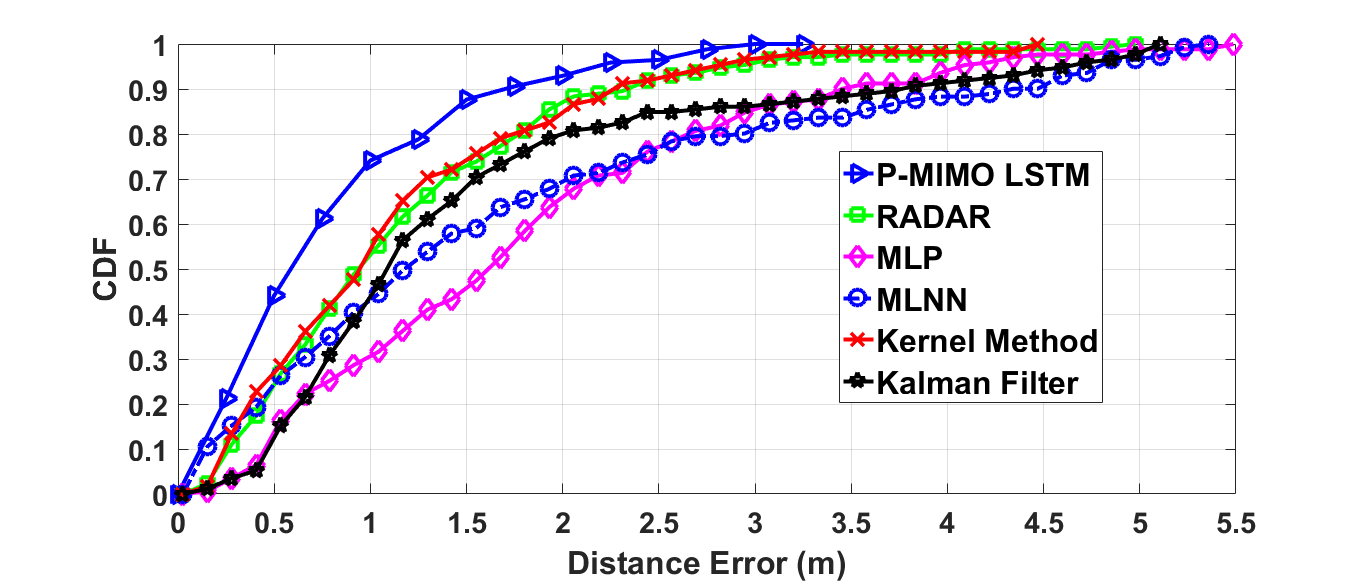}
\caption{The CDF of the localization error of P-MIMO LSTM and the other methods in literature.}
\label{fig:rnn_lstm}
\end{figure}

Fig.~\ref{fig:CDF_compare1}(d) compares the performance between vanilla RNN~\cite{Ho2017}, LSTM~\cite{Hochreiter1997}, GRU~\cite{Cho2014}, BiRNN, BiLSTM~\cite{Schuster1997} and BiGRU~\cite{Zhao2018}. All of the settings follow Table~\ref{table:setup}. Although the gap between these systems are close, LSTM still consistently has the best performance with the average error at $\mathrm{0.75{\pm}0.64~m}$ compared to $\mathrm{1.05{\pm}0.77~m}$ of RNN, $\mathrm{0.80{\pm}0.70~m}$ of GRU, $\mathrm{0.95{\pm}0.77~m}$ of BiRNN, $\mathrm{0.89{\pm}0.75~m}$ of BiLSTM and $\mathrm{0.93{\pm}0.81~m}$ of BiGRU, respectively. While $\mathrm{80\%}$ of the errors of RNN and BiLSTM are all within 1.5~m, the ones of LSTM and GRU are within 1.2~m. Some explanations are as follows.

Regarding vanilla RNN, there is a disadvantage of vanishing gradient at large $T$~\cite{Chung2014}. Therefore, RNN has difficulty to learn from the long-term dependency, i.e, $T=10$ in our case. Unlike the traditional RNN, both LSTM and GRU are able to decide whether to keep the existing memory from the past by their gates, i.e, forget gate in LSTM and reset gate in GRU. Intuitively, their performances in our experiment are both better than vanilla RNN because if LSTM and GRU detect an important feature from an input sequence at early stage, they easily carry this information over a long distance and capture potential long-distance dependencies. Furthermore, GRU is a simpler version of LSTM, which means that some of the feature of LSTM are reduced in GRU, e.g., the exposure of the memory content control~\cite{Chung2014}. Therefore, LSTM has a slightly better performance than GRU. For the bidirectional models including BiRNN, BiLSTM, BiGRU, they use both historical and future information, i.e., the initial location and the last location to predict the current location. In this paper, we assume only the very first location of each trajectory is known. As the ground truth of the last location is not incorporated, error is introduced into the bidirectional models. Therefore, the bidirectional models are not favorable in our work.

\subsection{Literature Comparison}
Fig.~\ref{fig:rnn_lstm} compares the proposed P-MIMO LSTM with the feedforward neural network MLP~\cite{Battiti2002}, multi-layer neural network (MLNN)~\cite{Dai2016} and the other conventional methods including KNN-RADAR~\cite{Bahl2000}, probabilistic Kernel method~\cite{Kushki2007} and Kalman filter~\cite{Au2013}. In our experiments, we reproduced the results of MLP and MLNN~\cite{Battiti2002,Dai2016}. The MLP model has three layers with only one 500-neuron hidden layer. In contrast, MLNN has five layers, i.e., 1 input layer, 3 hidden layers with 200, 200 and 100 neurons respectively and 1 output layer. The input of these memoryless methods is a single RSSI vector (11 RSSI readings) of a specific location, the output is a single location. The input of  P-MIMO is a trajectory with $T$ RSSI vectors ($11{\times}T$ RSSI readings) from $T$ time steps, the output is a trajectory including $T$ output locations.  P-MIMO clearly outperforms MLP with the maximum error of 3.4~m compared with 5.5~m of MLP.  $\mathrm{80\%}$ of LSTM model errors are within 1.2~m, which is much lower than 2.7~m of MLP. The maximum errors of conventional methods such as RADAR, Kalman filter and Kernel method are more significant 4.8~m, 5.0~m and 4.50~m, respectively. Besides, $\mathrm{80\%}$ of the errors of those methods are all within 2~m, 1.7~times higher than the proposed LSTM model. Table~\ref{table:Error1} lists the average errors between LSTM models and the other mentioned methods. Clearly, the accuracy of P-MIMO with 0.75~m dominates the other conventional feedforward neural networks with 1.65~m of MLNN~\cite{Dai2016} and 1.72~m of MLP~\cite{Battiti2002}. 

\subsection{Further Discussion}
\begin{figure}[!t]
\centering
\includegraphics[width=0.5\textwidth]{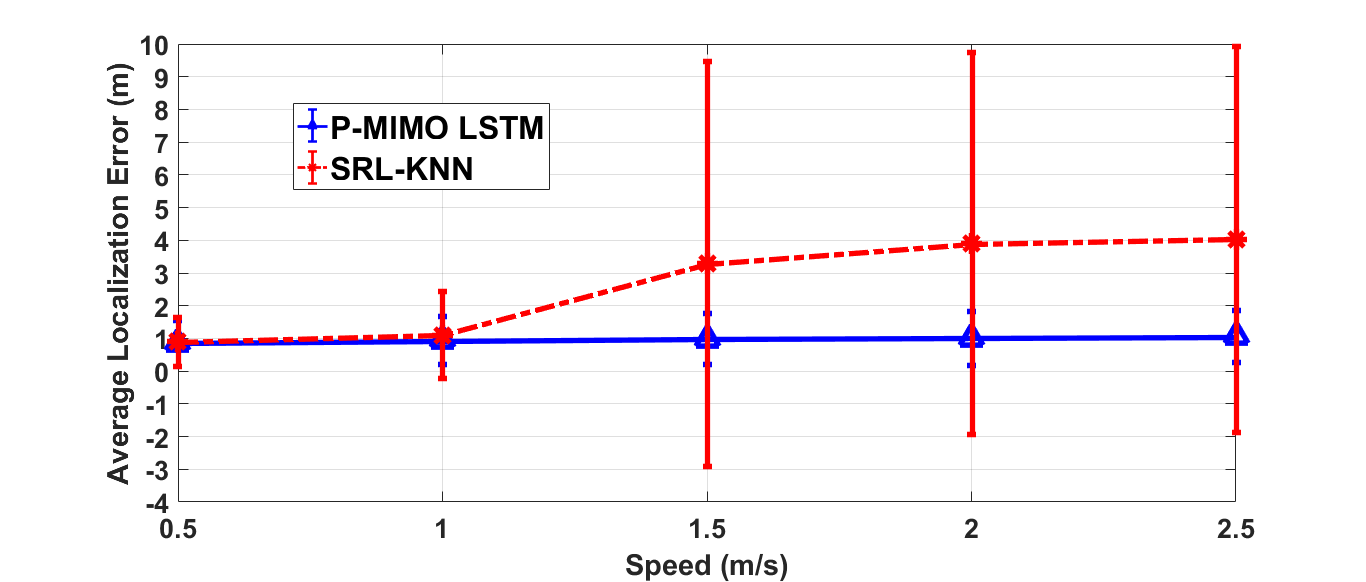}
\caption{Average localization errors with the error bars of P-MIMO LSTM and SRL-KNN in changing speed scenarios}
\label{fig:diff_speed}
\end{figure}

\subsubsection{Three Challenges of WiFi Indoor Localization}
As mentioned at the beginning, the proposed LSTM models can address three challenges of WiFi indoor localization.  Subsection~\ref{sub:mem_exp} has demonstrated that LSTM adopts sequential measurements from several locations in the trajectory and decreases  the spatial ambiguity significantly. In addition, RSSI instability and RSSI short collecting time per location  create the diverse values of RSSI readings in one location, which can lead to more locations having similar fingerprint distances. The proposed LSTM models can effectively remove the number of false locations which are far from the previous points based on the series of the previous measurements and predictions. Therefore, the adverse effects of the second and third challenge can be mitigated.  

\subsubsection{Training Time Requirement}
 \begin{table*}
\centering         
\caption{Average Localization Errors Of UJIIndoorLoc Database} \label{table:AverageErr2} 
\begin{tabular}{l c c c c c c} 
\hline           
 & \textbf{P-MIMO LSTM} & \textbf{RADAR \cite{Bahl2000}} & \textbf{MLNN \cite{Dai2016}} & \textbf{Kalman  Filter\cite{Au2013}} &  \textbf{MLP \cite{Battiti2002}} \\ 
Building 0 (m) & 4.5 $\pm$ 2.7 &  7.9 $\pm$ 4.9 & 7.6 $\pm$ 4.2  & 8.2 $\pm$ 5.0 & 9.2 $\pm$ 5.8\\
Building 1 (m) & 4.0 $\pm$ 3.8 &  8.2 $\pm$ 4.9 & 7.5  $\pm$ 3.3 & 8.4 $\pm$ 3.9 & 7.4 $\pm$ 4.4\\ 
All buildings (m)& 4.2 $\pm$ 3.2 & 8.1  $\pm$ 4.9 & 7.5  $\pm$ 3.8  & 8.2  $\pm$ 4.7 & 8.2  $\pm$ 5.2\\ 
\hline   
\end{tabular} 
\end{table*}
Compared with other conventional methods, e.g., KNN-RADAR~\cite{Bahl2000}, probabilistic Kernel method~\cite{Kushki2007} and Kalman filter~\cite{Au2013}, the proposed RNN method outperforms. However, those conventional methods do not require the training phase, which compares the current RSSI measurement with the ones in the database to get user locations directly. In contrast, the proposed RNN methods require to train the neural network model beforehand. The learning curve can determine the minimum required training time for the proposed RNN models. From Subsection~\ref{sub:training_require}, the optimal training trajectory samples is $\mathrm{10^4}$ and the optimal number of epochs is approximately 1,000 according to a 1 fold test. In the experiment, the training is executed on a home-built computer with an AMD FX(tm)-8120 Eight-Core CPU and an Nvidia GTX 1050 GPU. The running time is approximately 4~s per epoch. Therefore, the training time is approximately $\mathrm{4{\times}1000 = 4000~s}$ ($\cong$ 1 hour and 6 minutes).

In practical scenarios, more APs are used, more RSSI features can be extracted and better performance can be achieved. However, increasing the number of APs also creates more computational cost and extends the training time. Ref.~\cite{Dong2017} suggests to use LSTM with projection layer (LSTMP) to reduce the computational cost.  Furthermore, LSTMP was reported helpful to error rate reduction because parameter reduction helps the LSTM generalization. In the future work, when the number of APs are increased to achieve better accuracy, the idea of LSTMP can be applied to reduce the training time of our RNN models. 

\subsubsection{Impacts from speed variation}
Some conventional short-memory methods such as Kalman filter~\cite{Au2013}, SRL-KNN~\cite{Minh2018} have the constraints on the speed of the users. If a user changes moving speed rapidly, the localization accuracy of these methods will be degraded severely. In contrast, the proposed LSTM network is trained with random trajectories as described in Subsection~\ref{sub:map_gen} without strict constraints on the speed of the users. Fig.~\ref{fig:diff_speed} illustrates the average errors of the proposed P-MIMO LSTM and SRL-KNN using RSSI mean database with parameter $\mathrm{\sigma = 2~m}$~\cite{Minh2018}. The number of testing points are 344 locations following the backward and forward trajectory like Fig.~\ref{fig:floor_map}. The maximum speed is the instant speed of the robot between 2 random consecutive testing locations in a sampling time interval $\Delta{t}$. The number of locations having the maximum speed are $\mathrm{50\%}$ of the total testing points (172 locations). The rest of the locations have a random speed smaller than the maximum speed. When the maximum speed increases from 0.5~m/s to 2.5~m/s, the average errors of LSTM model stays stable around $\mathrm{0.85{\pm}0.75~m}$. On the other hand, SRL-KNN starts from the comparable result as P-MIMO LSTM with 0.90~m in the case of 0.5~m/s. After the maximum speed increases to above 1.5~m/s, the accumulated errors appear and the accuracy of SRL-KNN is significantly degraded to above 3~m with a large variation more than 5~m.     

\subsubsection{Impacts from Different Time Slots}
\begin{figure}[!t]
\centering
\includegraphics[width=0.5\textwidth]{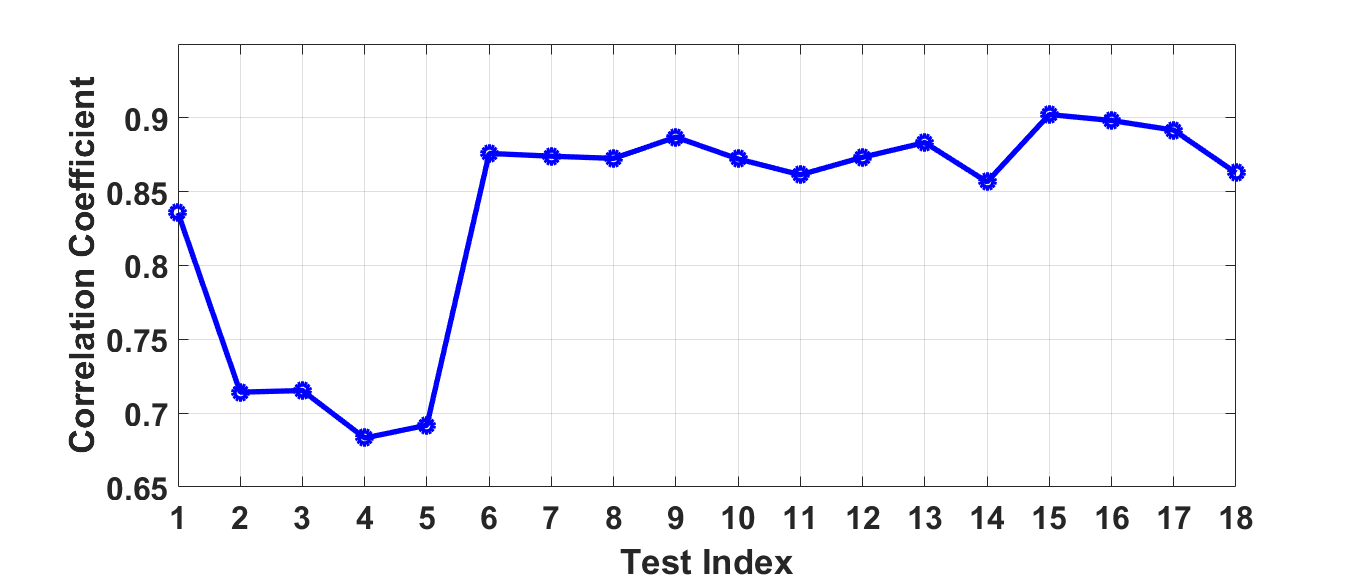}
\caption{Average correlation coefficient between different time trajectory tests and the database.}
\label{fig:corr_graph}
\end{figure}

The proposed RNN methods first learn the RSSI range characteristics of the environment offline in a prior training phase before using the learned characteristics during the testing phase. Therefore, the initial data in the training phase might not have the same distribution with the data in the testing phase \cite{Guo2019}. In our experiment, we address this problem with the support of our autonomous robot as shown in Fig.~\ref{fig:map_gen}(a). The robot is programmed to repeatedly navigate around the experimental area and collect the new data to update the database in different hours and days. The proposed RNN networks are trained with a wide variety of data reflecting the different characteristics of the environment at different time periods. Fig.~\ref{fig:corr_graph} illustrates the Pearson correlation coefficients between different time trajectory tests and the appropriate neighbour locations in our database. We repeatedly collect the testing trajectory with 175 locations as shown in Fig.~\ref{fig:floor_map}(a) at 18 random hours. The correlation coefficients range from 0.7 to 0.9, which proves that we can always find similarly distributed data in the database corresponding to each single test.

 Furthermore, according to the survey~\cite{Wu2018}, the percentage of stationary time when a user is not moving can exceed $\mathrm{80\%}$ for most mobile users. Some of the locations falling in the stationary period can serve as the anchor points to re-calibrate the network in the testing phase~\cite{Ramiro2017}. On the other hand, if other sensors such as camera are available, the additional data can help to increase the estimation accuracy of locations in the trajectory~\cite{Ramiro2017}. A detailed study of these issues is out of the scope of this paper but will be addressed in our future work.  

\subsubsection{Stability and Robustness}
\begin{figure}[!t]
\centering
\includegraphics[width=0.5\textwidth]{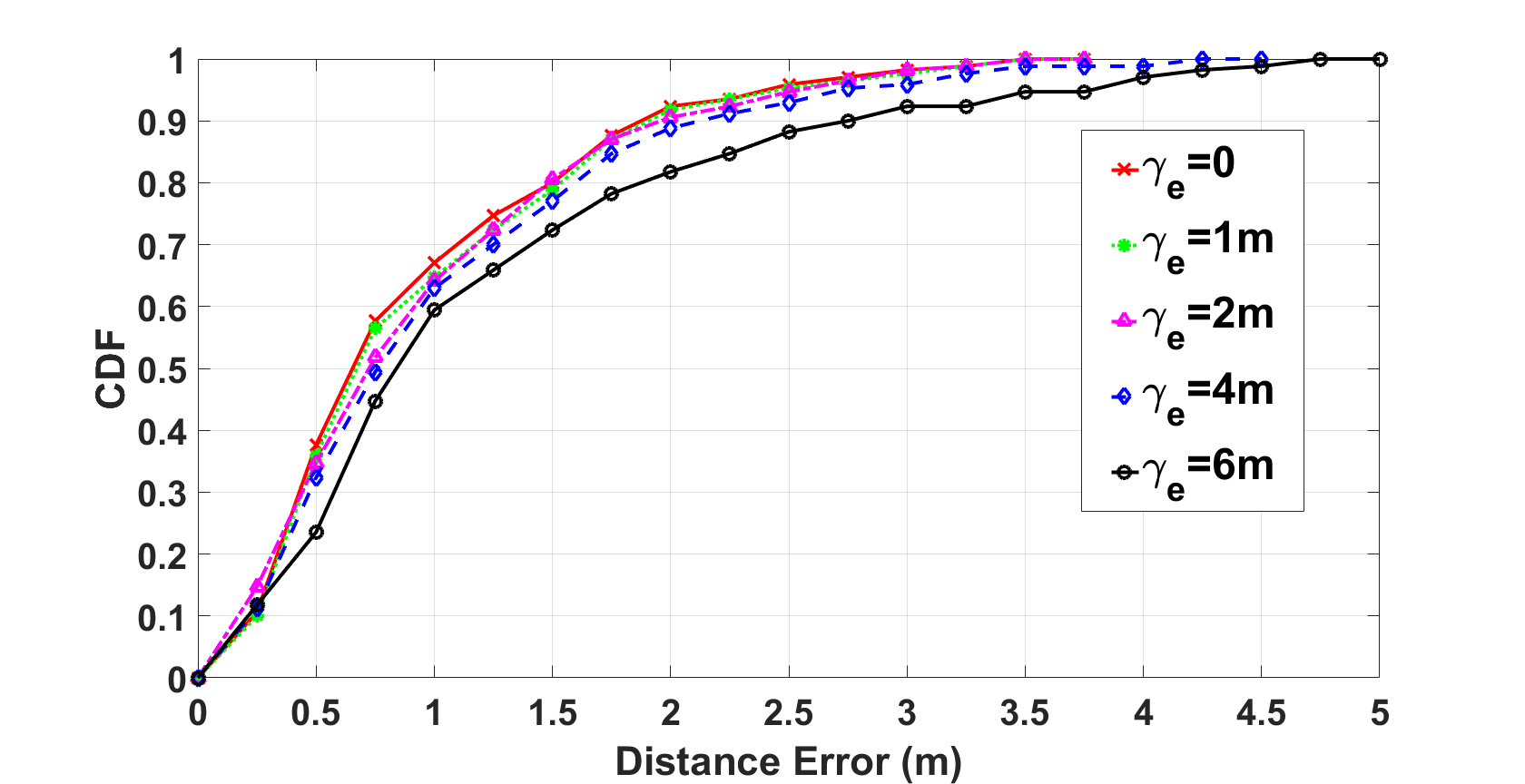}
\caption{CDF of P-MIMO LSTM localization errors  in different  historical data error scenarios.}
\label{fig:error_analysis}
\end{figure}
Since P-MIMO LSTM leverages the information of a user's previous positions to estimate the current location, the stability of P-MIMO LSTM depends on the accuracy of historical data from the previous steps. In order to investigate the propagation error due to the imperfect prior location estimation, Fig.~\ref{fig:error_analysis} illustrates the localization errors of P-MIMO LSTM with both the ideal and erroneous history data. Starting with the perfect historical coordinate $\bm{h}(x,y)$ for every location in the testing trajectory as illustrated in Section~\ref{sec:database}, an amount of Gaussian error is added to $\bm{h}$. The erroneous prior location $\bm{h^{\prime}}(x^{\prime}, y^{\prime})$ is obtained as: $ x^{\prime} = x + x_{e} \, , \, y^{\prime} = y + y_{e}$, where $x_e$ and $y_e$ are random variables that follow Gaussian distribution
\[ x_e \sim \mathcal{N}(0,\sigma_{x_{e}}^{2}) \, ; \, y_e \sim \mathcal{N}(0,\sigma_{y_{e}}^{2}) \, ; \, \gamma_{e} = \sqrt{\sigma_{x_{e}}^{2}+\sigma_{y_{e}}^{2}} \] 
Fig.~\ref{fig:error_analysis} shows that if the standard deviation error $\gamma_e$ of the historical data is within $2$ m, the localization accuracy is mostly similar to the ideal case, with a maximum error of $3.5$ m and $80\%$ of the error is $1.5$ m. When $\gamma_e$ increases to $4$ m and $6$ m, the accuracy becomes slightly worse with the maximum errors being around $5$ m and $80\%$ errors being around $1.80$ m and $2.3$ m, respectively. As shown in Table~\ref{table:Error1}, the average errors of P-MIMO LSTM is within $1$ m, i.e., $0.75 \pm 0.64$ m, which indicates that our proposed P-MIMO LSTM is robust to the localization error of the previous positions.

\subsection{Other Database Comparison}
\begin{figure}[!t]
\centering
\includegraphics[width=0.52\textwidth]{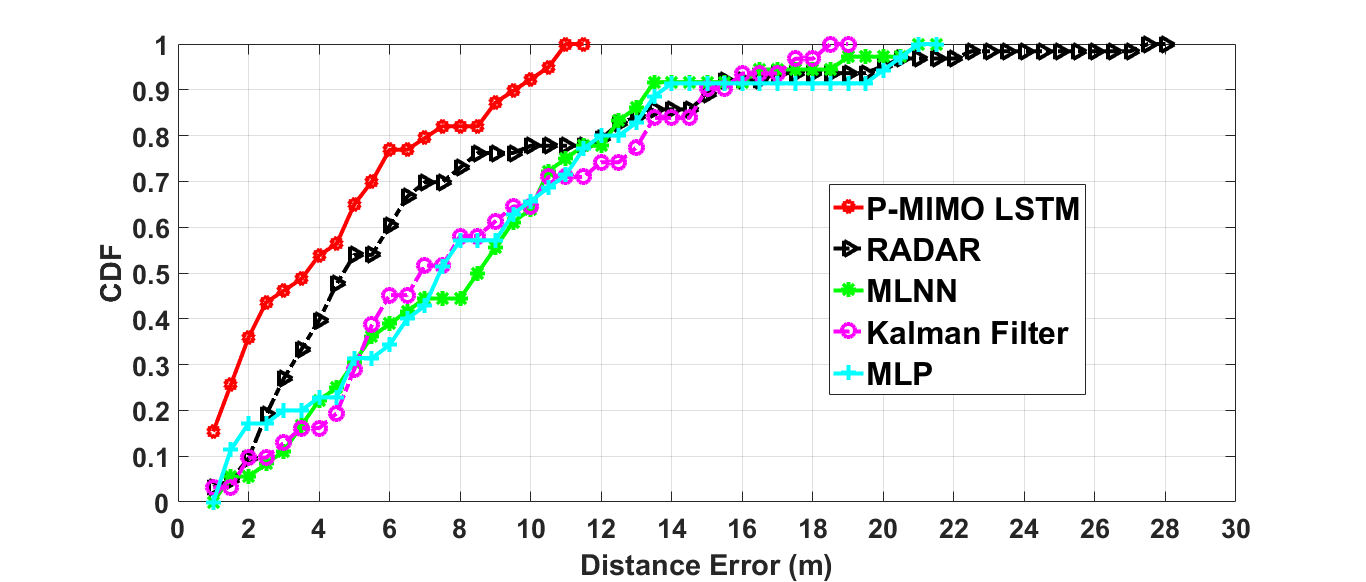}
\caption{Localization error CDF of UJIndoorLoc database for all buildings}
\label{fig:UJIndoor}
\end{figure}

The consistent effectiveness of the proposed LSTM system is proved by the published dataset, UJIIndoorLoc~\cite{Torres-Sospedra2014}. The reported average localization error~\cite{Torres-Sospedra2014} is 7.9~m. The database from 2 random phone users (Phone Id: 13, 14) in 2 different buildings (Building ID: 0 and 1)  are used to implement P-MIMO LSTM. Note that the grid size of UJIIndoorLoc is different from the collected database which affects the average localization error. However, the relative accuracy comparison between the proposed LSTM and conventional KNN, e.g., RADAR~\cite{Bahl2000} and Kalman  filter~\cite{Au2013} or feedforward neural network, e.g, MLP~\cite{Battiti2002} and MLNN~\cite{Dai2016} can verify the effectiveness of our algorithm. Table~\ref{table:AverageErr2} shows the average errors in meter of P-MIMO LSTM, RADAR, Kalman filter, MLP and MLNN for each separate building and for all 2 buildings in general. For all 2 buildings, the average error of P-MIMO LSTM is $\mathrm{4.2{\pm}3.2~m}$, significantly lower than the result of RADAR $\mathrm{8.1{\pm}4.9~m}$, MLNN $\mathrm{7.5{\pm}3.8~m}$, Kalman filter $\mathrm{8.2{\pm}4.7~m}$ and MLP $\mathrm{8.2{\pm}5.2~m}$. Furthermore, Fig.~\ref{fig:UJIndoor} compares the CDF of localization errors between those methods. In total, a 11.5~m maximum localization error is recorded for P-MIMO LSTM, 22~m for MLNN and the largest maximum localization error of 28~m for RADAR. Besides, $\mathrm{80\%}$ of the error is below 7~m in the case of P-MIMO LSTM, which is much lower than 12~m in the case of MLP, MLNN and RADAR.    

\section{Conclusions} \label{sec:conclude}
In conclusion, we have proposed recurrent neural networks for WiFi fingerprinting indoor localization. Our RNN solution has considered the relation between a series of the RSSI measurements and determines the user's moving path as one problem. Experimental results have consistently demonstrated that our LSTM structure achieves an average localization error of 0.75~m with $\mathrm{80\%}$ of the errors under 1~m, which outperforms  feedforward neural network, conventional methods such as KNN, Kalman filter and probabilistic methods. Furthermore, main challenges of those conventional methods including the spatial ambiguity, RSSI instability and the RSSI short collecting time have been effectively mitigated. In addition, the analysis of vanilla RNN, LSTM, GRU, BiRNN, BiLSTM and BiGRU with important parameters such as loss function, memory length, input and output features have been discussed in detail.  

\bibliographystyle{IEEEtran}
\bibliography{LSTM_paper_ref}

\end{document}